\documentclass[aps,prb,twocolumn, showpacs,superscriptaddress,citeauthorscript]{revtex4-2}

\usepackage{graphicx}  
\usepackage{dcolumn}   
\usepackage{bm}        
\usepackage{amssymb}   
\usepackage{amsmath}
\usepackage{gensymb}
\usepackage{color}  

\usepackage[normalem]{ulem}

\usepackage{hyperref}

\usepackage[dvipsnames]{xcolor}
\setcitestyle{super,open={},close={}}
\usepackage[normalem]{ulem}

\hypersetup{
    colorlinks,
    linkcolor={blue!50!black},
    citecolor={blue!50!black},
    urlcolor={blue!80!black}
}

\begin{document}

\title{Parametric longitudinal coupling of a semiconductor charge qubit and a RF resonator}

\author{V. Champain}
\email{E-mail: victor.champain@cea.fr}
\author{S. Zihlmann}
\email{E-mail: simon.zihlmann@cea.fr}
\affiliation{Univ. Grenoble Alpes, CEA, Grenoble INP, IRIG-Pheliqs, Grenoble, France.}
\author{A. Chessari}
\affiliation{Univ. Grenoble Alpes, CEA, IRIG-MEM-LSim, Grenoble, France.}
\author{B. Bertrand}
\author{H. Niebojewski}
\affiliation{Univ. Grenoble Alpes, CEA, LETI, Minatec Campus, Grenoble, France.}
\author{\'{E}. Dumur}
\author{X. Jehl}
\author{V. Schmitt}
\author{B. Brun}
\author{C. Winkelmann}
\affiliation{Univ. Grenoble Alpes, CEA, Grenoble INP, IRIG-Pheliqs, Grenoble, France.}
\author{Y.M. Niquet}
\author{M. Filippone}
\affiliation{Univ. Grenoble Alpes, CEA, IRIG-MEM-LSim, Grenoble, France.}
\author{S. De Franceschi}
\author{R. Maurand}
\affiliation{Univ. Grenoble Alpes, CEA, Grenoble INP, IRIG-Pheliqs, Grenoble, France.}

\date{\today}

\begin{abstract}
In this study, we provide a full experimental characterization of the parametric longitudinal coupling between a CMOS charge qubit and an off-chip RF resonator. Following Corrigan {\it et. al.}, Phys. Rev. Applied 20, 064005 (2023), we activate parametric longitudinal coupling by driving the charge qubit at the resonator frequency. Managing the crosstalk between the drive applied to the qubit and the resonator allows for the systematic study of the dependence of the longitudinal and dispersive charge-photon couplings on the qubit-resonator detuning and the applied drive. Our experimental estimations of the charge-photon couplings are perfectly reproduced by theoretical simple formulas, without relying on any fitting parameter. We go further by showing a parametric displacement of the resonator's steady state, conditional on the qubit state, and the insensitivity of the longitudinal coupling constant on the photon population of the resonator. Our results open to the exploration of the photon-mediated longitudinal readout and coupling of multiple and distant spins, with long coherent times, in hybrid CMOS cQED architectures.

\end{abstract}

\maketitle
The fast and efficient measurement and manipulation of electron or hole spins in semiconducting materials, such as silicon or germanium, is a key challenge to devise scalable spin-qubit architectures. While fidelities above $99\,\%$ have been demonstrated for single and two qubit gates \cite{Xue_Nature_2022,Noiri_Nature_2022,mills2021}, state preparation and measurement (SPAM) errors are still generally around $10\,\%$. Due to the minute signal originating from a single spin, the measurement of spin-qubit states usually involves spin-to-charge conversion followed by charge detection via nearby electrometers \cite{hanson_spins_2007,elzerman_single-shot_2004,petta_coherent_2005-1}. The inclusion of such charge sensors complicates the design of scalable quantum processors \cite{Borsoi2023}. Alternative options stem from the concept of circuit quantum electrodynamics\cite{blais_circuit_2021}(cQED). Such hybrid cQED platforms require connecting semiconductor quantum dots (QDs) to on- or off-chip microwave resonators \cite{childress_mesoscopic_2004,burkard_ultra-long-distance_2006,hu_strong_2012,jin_strong_2012,frey_dipole_2012,petersson_circuit_2012,viennot_coherent_2015,samkharadze_strong_2018,mi_coherent_2018,clerk_hybrid_2020,yu_strong_2023}.

Most of these implementations rely on the dipole coupling between the qubit and the electric field of the resonator, resulting in a transverse interaction of the form $\sigma_x(a+a^\dagger)$, where $\sigma_i$ represents the qubit Pauli matrices and $a$ is the bosonic annihilation operator of the resonator. Widely used in cQED, this transverse interaction is fundamental for qubit state readout in the dispersive regime when the qubit and cavity frequencies are detuned \cite{blais_circuit_2021}.

In QD qubits, the transverse interaction has been employed to demonstrate strong charge- or spin-photon coupling\cite{hu_strong_2012,jin_strong_2012,frey_dipole_2012,petersson_circuit_2012,viennot_coherent_2015,samkharadze_strong_2018,mi_coherent_2018,yu_strong_2023}, as well as distant spin-spin interaction mediated by photons \cite{borjans_resonant_2020,harvey-collard_coherent_2022,Dijkema_Two_2023}. However, this approach can also introduce perturbations, such as qubit decay due to the Purcell effect \cite{houck_controlling_2008}, or alter qubit dynamics \cite{bertet_dephasing_2005,schuster_ac_2005,gambetta_qubit-photon_2006}. Consequently, there is growing interest in coupling schemes that mitigate these limitations \cite{blais_circuit_2021,kerman_quantum_2013}.

One promising approach to perform faster readout involves the use of the longitudinal interaction of the form $\sigma_z (a+a^\dagger)$. When combined with parametric modulation, the longitudinal interaction produces a qubit-dependent resonator displacement, enabling fast and quantum non-demolition (QND) qubit measurement, as recently demonstrated for transmon qubits \cite{didier_fast_2015,touzard_gated_2019,ikonen_qubit_2019}. Such a coupling would also facilitate qubit-qubit interaction in the form of non-resonant ZZ couplings between distant spins connected through a shared cavity \cite{royer_fast_2017,michal_tunable_2023,harvey_coupling_2018,bottcher_parametric_2022}.

Parametric driving of the energy detuning axis of a double QD charge qubit is predicted to induce a qubit cavity longitudinal interaction\cite{kerman_quantum_2013,beaudoin_coupling_2016,ruskov_quantum-limited_2019,ruskov_modulated_2021,ruskov_longitudinal_2023}. Recently, this idea was experimentally applied to read out a charge qubit in a Si/SiGe QDs device \cite{corrigan_longitudinal_2023,harpt_ultra-dispersive_2024}, utilizing a parametric drive on one of the gates defining the QD.

However, longitudinal and dispersive couplings are competing mechanisms~\cite{ikonen_qubit_2019}, which may be difficult to discriminate  due to  crosstalks causing undesirable photon leakage into the resonator. In this work, we take advantage of a triple QD device to experimentally investigate this issue. We then turn to a double QD configuration (single charge qubit) and demonstrate an experimental protocol to probe the longitudinal interaction, based on a compensation of crosstalk. Our measurements of the longitudinal and dispersive coupling are reproduced by simple theoretical expressions~\cite{ikonen_qubit_2019,ruskov_modulated_2021,corrigan_longitudinal_2023,chessari_unifying_2024}, without any fitting parameter. In particular, we demonstrate the expected proportionality between the longitudinal coupling strength and the parametric drive amplitude as well as its dependence on the gate-tunable qubit energy. In addition, we reveal two important characteristics of the longitudinal readout mechanism. First, we verify that it does not dependent on the photon population in the resonator. Second, we show that the qubit-state dependent cavity displacement is qualitatively and quantitatively different from the one observed in transverse dispersive readout, in agreement with theoretical expectation.

\section{Schemes to read-out a triple quantum-dot system}\label{sec:crosstalk}

\begin{figure}
    \centering
    \includegraphics[width = \linewidth]{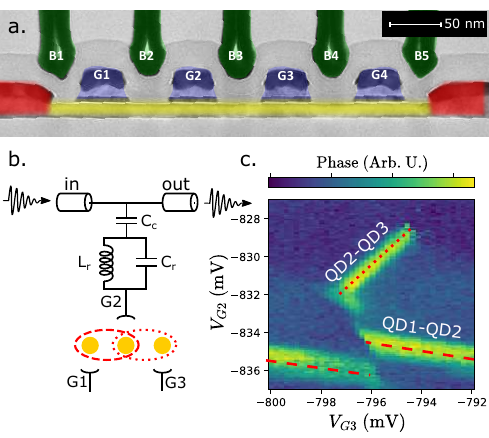}
    \caption{\textbf{Device and characterization. a.} False-colored transmission electron micrograph of the device. Yellow: Silicon nanowire. Red: Boron-doped leads, acting as reservoirs for holes. Blue: four parallel gates to define quantum dots by accumulation. Green: five parallel gate to tune the potential barrier between the dots. \textbf{b.} Circuit model of the system. A LC tank-circuit is galvanically connected to gate G2 and capacitively coupled to a transmission line. Three dots are schematically represented below gates G1, G2 and G3. Note that the DC circuit elements are omitted for clarity. \textbf{c.} Phase of the signal transmitted through the transmission line as a function of the voltages on G2 and G3.The red-dotted (resp. dashed) line corresponds to interdot transitions between QD2 and QD3 (resp. QD1 and QD2). }
    \label{fig1}
\end{figure}

The metal-oxide-semiconductor (MOS) device studied in this work consists of a silicon-on-insulator nanowire with a $50 \times 11$ nm$^2$ rectangular cross-section. The nanowire is covered by a first layer of 4 parallel plunger gates (G1 to G4) with a pitch of 80 nm, confining QDs, and by a second layer of 5 barrier gates (B1 to B5), tuning the tunnel coupling between adjacent dots\cite{bedecarrats_new_2021}, see Fig.~\ref{fig1}a. With p-doped leads, holes are accumulated under negatively-biased gates. 
A LC resonator of frequency $\omega_r/2\pi = 526$ MHz and decay rate $\kappa/2\pi = 13.9$ MHz is connected to gate G2 and capacitively coupled to a transmission line for readout. The RF signal passing through the feed-line is demodulated via homodyne detection. In the remainder of the manuscript, the readout tones are always at frequency $\omega_r/2\pi$. 

\begin{figure*}
    \centering
    \includegraphics[width = \linewidth]{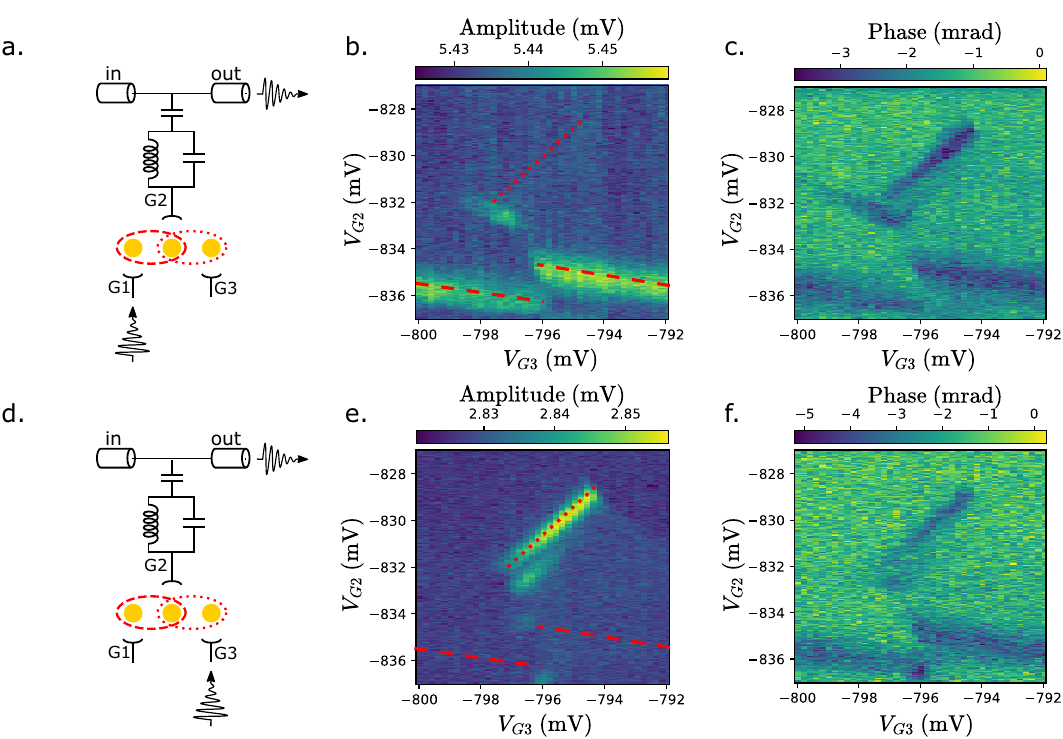}
    \caption{\textbf{Longitudinal measurement by driving on the gates. a.} (resp. \textbf{d.})  The readout tone is applied on gate G1 (resp. G3). \textbf{b.} (resp. \textbf{e.}) Amplitude of the signal measured at the output of the transmission line, demodulated at frequency $\omega_r/2\pi$, with the drive on gate G1 (resp. G3). \textbf{c.} (resp. \textbf{f.}) Phase shift of the transmitted signals.}
    \label{fig2}
\end{figure*}

To directly showcase the difference and interplay between transverse and longitudinal readout mechanisms, we consider a triple QD (QD1, QD2, and QD3) formed by hole accumulation below three plunger gates (G1, G2 and G3, respectively). First, we use standard dispersive readout based on gate reflectometry \cite{colless_dispersive_2013} to measure the stability diagram of the system as a function of $V_{\rm{G2}}$ and  $V_{\rm{G3}}$.  Figure \ref{fig1}b presents the driving protocol which consists in measuring the transmitted signal through the feed-line (from in to out).  The stability diagram obtained from the phase of the transmitted signal is shown in Fig.~\ref{fig1}c. In dispersive readout\footnote{Notice that actually the phase signal always arises from the dispersive shift $\chi$, which coincides with the quantum capacitance in the adiabatic limit where $\omega_r\ll\omega_q$, see also discussion in Refs.~\cite{park_from_2020,chessari_unifying_2024}.}, the phase signal arises from the quantum capacitance associated to charge transitions between tunnel-coupled QDs \cite{colless_dispersive_2013,gonzalez-zalba_probing_2015,ilani_measurement_2006,ota_wide-band_2010,mizuta_quantum_2017}.  Here, we notice two types of transitions with different slopes. The positive-slope ridge (highlighted by a dotted line) corresponds to a charge tunneling between QD2 and QD3, while the negative-slope ridges (highlighted by dashed lines) correspond to charge tunneling between QD1 and QD2. In a cQED framework, the phase signal at each interdot transition originates from quantum interaction between the charge states of double-QD qubit and the photonic modes of the LC resonator. With a charge qubit energy above the resonator frequency, the dispersive interaction decreases the resonator frequency at the interdot transition, giving rise to the phase signal seen in Fig.~\ref{fig1}c \cite{petersson_charge_2010,colless_dispersive_2013,penfold-fitch_microwave_2017,ezzouch_dispersively_2021}.

In the following, we measure the same stability diagram as in Fig.~\ref{fig1}c, but using readout protocols based on the parametric driving of either the QD1/QD2 or the QD2/QD3 charge transition \cite{corrigan_longitudinal_2023}. This is achieved by applying a RF tone at frequency $\omega_r/2\pi$ either on G1 or on G3, as schematically shown in Fig.~\ref{fig2}a and \ref{fig2}d. Importantly, no signal is applied at the input of the feed line. The stability diagrams in amplitude and phase are presented in Figs.\ref{fig2}b,c and \ref{fig2}e,f for parametric driving applied to G1 and G3, respectively.

The stability diagrams in amplitude exhibit noticeable differences. Indeed, when parametrically driving G1 (G3), only the charge transition QD1/QD2 (QD2/QD3) is visible as shown in Fig.~\ref{fig2}b (\ref{fig2}e). The observed amplitude peak signals a parametrically driven enhancement of the photon population in the resonator. Note that the amplitude in the blockaded regions is non-zero. The stability diagrams in phase are, however, similar and reveal the same charge transitions highlighted in Fig.~\ref{fig1}c. These two observation are a clear indication of a residual crosstalk between G1 (G3) and the resonator.

However, such crosstalks are not sufficient to explain all the observations of Fig.~\ref{fig2}. Photon leakage into the resonator would in principle lead to dispersive readout, resulting in a ohase signal.  Figure~\ref{fig1}c clearly shows that the resonator is dispersively coupled to both charge qubits with comparable strength (\textit{i.e}. comparable lever arms and comparable interdot tunnel couplings). Thus, the presence of only specific signals in Fig.~\ref{fig2}b,e, associated with well identified interdot transitions, is rather consistent with the presence of the parametrically-modulated longitudinal coupling, originally observed by Corrigan {\it et al.}~\cite{corrigan_longitudinal_2023}. Giving an unambiguous characterization of such longitudinal coupling, unpolluted from residual crosstalk, is the aim of the next section.

\section{Quantitative analysis}

Following the above demonstration of parametric driving via gate-modulated longitudinal coupling as well as the simultaneous existence of residual dispersive coupling, we shall now turn to a quantitative study of these competing readout mechanisms. To this purpose, we shall restrict our attention to the prototypical case of a charge qubit formed by a single hole delocalized in double QD under gates G2 and G3 (see Appendix C for more information about the charge configuration). Beforehand we introduce the necessary theoretical framework.

\subsection{Theory}
\begin{figure}
    \centering
    \includegraphics[width=\linewidth]{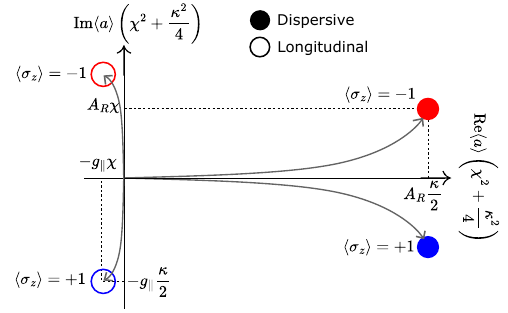}
    \caption{Resonator pointer states depending on the readout mechanism and the state of the qubit. Note that the dispersive points have been rotated by $\pi/2$ as compared to Eq.~\eqref{Eq:a_disp}. We highlight figurative transient regimes with grey arrows.}
    \label{fig3}
\end{figure}

A single charge trapped inside a double quantum dot has two localized states $|L\rangle$ and $|R\rangle$. The Hamiltonian of this two-level system is given by $H_c = \varepsilon/2 (|L\rangle \langle L| - |R\rangle \langle R|) + t (|L\rangle \langle R| + |R\rangle \langle L|) $, where $\varepsilon$ is the energy detuning between the two dots, and $t$ is the tunnel coupling. The double QD acts as an electric dipole that interacts with the oscillating electric field of a microwave resonator, described by bosonic operators $H_r =\omega_r a^\dagger a $ (with the convention $\hbar=1$). The photon-dipole interaction term reads $H_{int} = g_0 (|L\rangle \langle L| - |R\rangle \langle R|) (a+a^\dagger)$,  where $g_0$ is the charge-photon coupling. In the qubit basis, the Hamiltonian $H_c$ takes the diagonal form  $H_c = \Omega_q/2\, \sigma_z$ with $\Omega_q = \sqrt{\omega_q^2 + \varepsilon^2}$ and $\omega_q = 2t$ being the charge qubit energy at zero detuning. In the same  basis, the dipole interaction results in the sum of a transverse and a longitudinal term:
\begin{equation}
    H_{int} = \frac{\omega_q g_0 }{ \Omega_q} \sigma_x ( a+a^\dagger) + \frac{\varepsilon g_0 }{ \Omega_q} \sigma_z (a+a^\dagger).
\end{equation} 
This expression explicitly shows the presence of a {\it static} longitudinal coupling between the qubit and the resonator at finite detuning ($\varepsilon\neq0$). This coupling dominates at large detuning ($\varepsilon \gg \omega_q$), where the transverse interaction is suppressed ($\omega_q / \Omega_q \ll 1$). However, this regime is not of interest since the qubit is largely susceptible to charge noise.

Thus, for the remainder of the manuscript, we shall set the double QD to $\varepsilon=0$, unless told otherwise. In this regime, the longitudinal interaction can be magnified by driving the qubit at the resonator frequency. The RF tone therefore modulates the detuning with amplitude $A_q$ and frequency $\omega_r/2\pi$ such that $\varepsilon = A_q \cos{\omega_r t}$. The charge-photon Hamiltonian can be re-written as~\cite{ikonen_qubit_2019,corrigan_longitudinal_2023,chessari_unifying_2024}:
\begin{equation}\label{Eq:ham_lon}
H_\parallel = \frac {\omega_q+\chi}2\sigma_z + (\omega_r + \chi \sigma_z) a^\dagger a   + g_\parallel\sigma_z (a+a^\dagger) \,,
\end{equation}
where $\chi$ stands for the dispersive coupling and $g_\parallel$ for the longitudinal coupling strength. If $A_q < \omega_q$ (weak-drive limit)  and $(\omega_q-\omega_r) \sim (\omega_q+\omega_r) \gg g_0 $ (dispersive regime), these coupling can be expressed as~\cite{ikonen_qubit_2019,chessari_unifying_2024}
\begin{align}
    \chi &= 2g_0^2 \frac{\omega_q}{\omega_q^2 - \omega_r^2} \label{Eq:chi}\,,\\
    g_\parallel &= \frac{g_0A_q}{2} \frac{\omega_q}{\omega_q^2-\omega_r^2} = \frac{\chi A_q}{4g_0}\,.\label{Eq:g_par}
\end{align}
These expression are consistent with those derived in Ref.~\cite{corrigan_longitudinal_2023} in the adiabatic limit ($A_q,\omega_r\ll t$) and at zero detuning ($\varepsilon=0$), where both the dispersive and longitudinal couplings become proportional to $1/\omega_q$~\cite{park_from_2020,chessari_unifying_2024}, that is the curvature of the qubit~\cite{ruskov_quantum-limited_2019,ruskov_modulated_2021}. 
Note that $\chi$ differs from the usual $g^2/(\omega_q - \omega_r)$  because we need to account for counter-rotating oscillations \cite{Zueco2009} when $\omega_r  \ll \omega_q$.
In our experiment, the resonator frequency, the decay rate, and the charge-photon coupling are fixed parameters, i.e.  $\omega_r/2\pi = 526$\,MHz and $\kappa/2\pi =13.9$ MHz, $g_0/2\pi = 25.0$ MHz (see Appendix~\ref{app:resonator}).

The last term on the right-hand side of Eq.~\eqref{Eq:ham_lon} embodies the fact that the resonator can be populated via a qubit-dependent ($\sigma_z$) longitudinal drive. Equation \eqref{Eq:ham_lon} results in a resonator steady-state field~\cite{didier_fast_2015,corrigan_longitudinal_2023}:
\begin{equation}
    \langle a \rangle_{\parallel} = - g_\parallel \frac{\chi +i \frac{\kappa}{2}\langle \sigma_z \rangle}{\chi^2  + \frac{\kappa^2}{4}} \,.\label{Eq:a_par}
\end{equation}
Figure \ref{fig3} presents schematically the time-evolution of the resonator field in phase space for the two qubit eigenstates ($\langle \sigma_z \rangle = \pm 1$) starting from the vacuum state. In practice, these pointer states can be resolved by homodyne detection to perform qubit measurement \cite{blais_cavity_2004}. For comparison, we also show the evolution of the resonator field expectation in the case of pure dispersive readout, where :
\begin{equation}
    \langle a \rangle_{\rm{disp}} = -i A_R \frac{\chi \langle \sigma_z \rangle +i \frac{\kappa}{2}}{\chi^2 + \frac{\kappa^2}{4}}\,.\label{Eq:a_disp}
\end{equation}
This expression is obtained from Eq.~\eqref{Eq:ham_lon} upon removing the parametric drive ($A_q=0$) and adding a classical drive directly to the resonator $2A_R\cos(\omega_rt)(a+a^\dagger)$.

\subsection{Measurement of the longitudinal coupling}

Based on Eq.~\eqref{Eq:a_par} and assuming the charge qubit to be in its ground state, the longitudinal coupling reads : 
\begin{equation}
    g_\parallel= |\langle a \rangle_{\parallel}| \sqrt{\chi^2 + \kappa^2 /4}\,,
    \label{Eq:g_par_exp}
\end{equation}
where $|\langle a \rangle_{\parallel}|$ can be deduced from the measurement of the output-field quadratures $I$ and $Q$ via a calibration procedure discussed in Appendix~\ref{app:resonator}. This provides a recipe to measure the longitudinal coupling. 

\begin{figure}
    \centering
    \includegraphics[width=\linewidth]{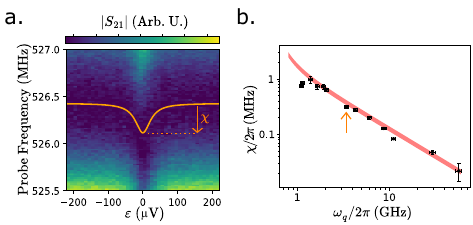}
    \caption{\textbf{Dispersive shift characterization.} \textbf{a.} Spectroscopic measurement of the resonator while varying the detuning $\varepsilon$ at $\omega_q/2\pi = 3.4$ GHz. $\chi$ is extracted as the difference between the resonant frequency of the LC resonator at $\varepsilon =0$  and $|\varepsilon| \gg \omega_q$, here $\chi = 313$ kHz \textbf{b.} Dispersive shift as a function of $\omega_q$. The red area shows the expected value using Eq.~\eqref{Eq:chi} and the estimated $g_0$ with its uncertainty, without any additional fitting parameters.}
    \label{fig4}
\end{figure}

To measure the dispersive shift $\chi$, we perform a spectroscopy of the LC resonator while varying the energy detuning in the double QD. Figure \ref{fig4}a presents the transmission $|S_{\rm{21}}|$ through the feed-line depending on the probe frequency and the detuning $\varepsilon$. The orange line overlaid on the data of Fig.~\ref{fig4}a highlights the resonance frequency and reveals the dispersive shift $\chi$ of the resonator interacting with the charge qubit \cite{frey_dipole_2012}. By varying the barrier gate voltage, we modify the tunnel rate between the two dots to change the charge qubit energy at zero detuning from $\omega_q/2\pi=1$\,GHz to $\omega_q/2\pi=57$\,GHz (see Appendix~\ref{app:charge}). The spectroscopy measurement of Fig.~\ref{fig4}a is repeated for each qubit energy to obtain the $\chi(\omega_q)$ dependence. The results are shown in Fig.~\ref{fig4}b, where the experimental data points are plotted together with the $\chi(\omega_q)$  curve calculated using Eq.~\eqref{Eq:chi} (see Appendix~\ref{app:resonator} and~\ref{app:charge}), without additional fitting parameters. In practice, we find that $\chi \ll \kappa$ for all $\omega_q$ values, implying that $\chi$ plays a minor role in the investigated experimental regime.

\begin{figure}
    \centering
    \includegraphics[width=\linewidth]{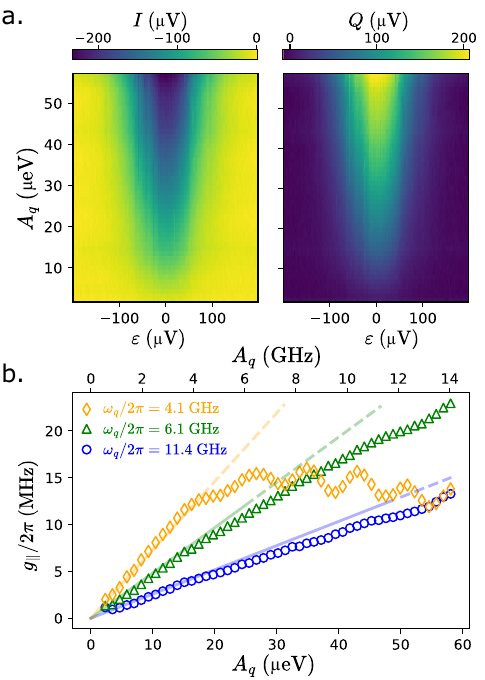}
    \caption{\textbf{Longitudinal coupling characterization. a.} Quadratures $I$ and $Q$ of the demodulated field while varying the drive amplitude on gate G3, and the electrical detuning for $\omega_q/2\pi = 6.1$ GHz. \textbf{b.} Extracted longitudinal coupling strength $g_\parallel$ as a function of the drive amplitude for three different qubit energy. The model (\ref{Eq:g_par}) is represented in solid lines up to $A_q = \omega_q$, without any additional fitting parameters. The model's lines become dashed for $A_q > \omega_q$. Note that the x-axis has been duplicated in frequency units above the graph.}
    \label{fig5}
\end{figure}

We can now proceed with the direct measurement of the longitudinal coupling $g_\parallel$. To do this, we perform the homodyne measurement of the $I$ and $Q$ components as a function of the $A_q$ drive, applied to G3. To perform a measurement of $g_\parallel$ unbiased from the residual crosstalk between the driving line, the resonator and the feed line, see Section~\ref{sec:crosstalk}, we apply a compensation tone to the feed-line input, as in Refs.\cite{ikonen_energy-efficient_2017,touzard_gated_2019}. The amplitude and phase of this tone are adjusted to cancel the output signal at large detuning ($ |\varepsilon | \gg \omega_q$), where charge is fully localized in one of the QDs  (see Appendix~\ref{app:cancellation}). In this regime, the dipole interaction between the charge qubit and the resonator becomes negligible thereby preventing parametric excitation of the photonic modes in the resonator. 

We thus measure $I$ and $Q$ as a function of $\varepsilon$ and $A_q$ while simultaneously applying the compensation tone. Representative results are shown in  Fig.~\ref{fig5}a for a qubit frequency  $\omega_q/2\pi = 6.1$ GHz. 
First, we note that both $I$ and $Q$ vanish at large detuning denoting a properly calibrated compensation tone.  Second, $I$ ($Q$) shows a dip (peak) centered at $\varepsilon=0$. The height of this dip (peak) increases almost linearly with $A_q$, indicating the parametric nature of the excitation. We repeat the same type of measurement for other qubit frequencies, {\it i.e.} by varying each time the interdot tunnel coupling. By converting the values of $I$ and $Q$ at zero detuning in $\langle a \rangle$ and using Eq.~\eqref{Eq:g_par_exp}, we can finally infer the longitudinal coupling strength as a function of the driving amplitude at three different qubit frequencies.

The results for the three qubit frequencies are shown in Fig.~\ref{fig5}b together with the corresponding linear-in-$A_q$ dependence expected from Eq.~\eqref{Eq:g_par}.  The remarkable agreement obtained  with no fitting parameter confirms the validity of the measurement protocol and demonstrates the realization of parametric longitudinal driving. We notice that,  since $\omega_q \gg \omega_r$,  the slope of the $g_\parallel(A_q)$ linear dependence is, to a good approximation, inversely proportional to the qubit frequency. Moreover, the linear increase of $g_\parallel$ is observed up to $A_q \sim \omega_q$, followed by a tendency to saturation possibly due to the breakdown of the weak-drive approximation $A_q < \omega_q$. 

\subsection{Influence of the photon population of the resonator}

\begin{figure}
    \centering
    \includegraphics[width=\linewidth]{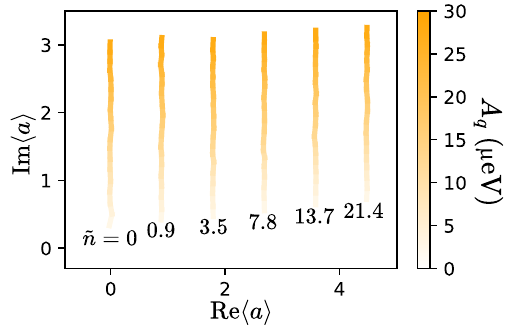}
    \caption{Resonator displacement field while increasing the drive amplitude for different initial added photon population  $\tilde{n}$. Here $\omega_q/2\pi = 4.1$ GHz. The trace at $\tilde{n} = 0 $ actually corresponds to the trajectory of ($I$,$Q$) as a function of $A_q$ at zero detuning $\varepsilon = 0$ from Fig.~\ref{fig5}a.}
    \label{fig6}
\end{figure}

In the measurement protocol discussed above,  the compensation tone yields $(I,Q) \sim (0,0)$ at large $\varepsilon$. This cancelling of the output signal does not necessarily imply a zero photon population in the resonator which could still be classically driven by a residual direct coupling to the $A_q$ drive.  
According to Eq.~\eqref{Eq:g_par}, however, the longitudinal coupling strength does not depend on the photon population. Hence the results shown in Fig.  \ref{fig5} should not be affected by a finite photon population superimposed to the one induced by the parametric driving via the charge qubit. We verify this statement by deliberately adding a drive of constant amplitude, on top of the compensation tone, at the input of the feed line. This additional drive creates an extra number of photons $\tilde{n}$ in the resonator, offsetting the $(I,Q)$ output signal. Figure \ref{fig6}  shows the results of homodyne detection at zero detuning for different values of $\tilde{n}$ (see Appendix~\ref{app:resonator} for the correspondence between $\tilde{n}$ and the amplitude of the additional drive). For a convenient representation, a global phase has been added to all measurements in such a way that the parametrically induced output signal lies along the imaginary axis.  
From Fig.~\ref{fig6}, we conclude that the measured displacement field is independent of the initial photon population.

\subsection{Qubit conditional displacement}

\begin{figure}
    \centering
    \includegraphics[width=\linewidth]{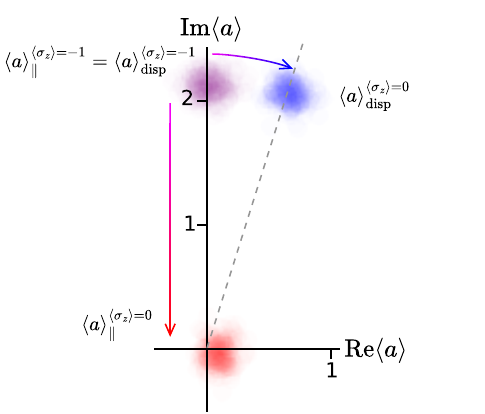}
    \caption{Resonator states measured for two different qubit states and two drive mechanisms. Here $\omega_q/2\pi = 4.1$ GHz and $A_q = 20$ µeV}
    \label{fig7}
\end{figure}

The parametric longitudinal term of the Hamiltonian in Eq.~\eqref{Eq:ham_lon} can be interpreted as a qubit-dependent conditional  displacement of the resonator state  \cite{touzard_gated_2019}. In all measurements discussed so far, the qubit was always in its ground state, \textit{i.e.} $\langle \sigma_z \rangle = -1$.  In the following, we apply an additional continuous drive of frequency $\omega_q/2\pi$ to investigate the effect of driving the qubit to an incoherent mixture $\langle \sigma_z \rangle \sim 0$. Figure \ref{fig7} shows the steady state pointer states in phase space measured by homodyne detection for the two qubit states (ground or mixed) for both readout schemes (dispersive and parametric longitudinal). To highlight the difference between longitudinal and dispersive readout, we operate on the phase and amplitudes of the drives $A_{q,R}$ in such a way that the signal sits on the same point in the  $(I,Q)$ when $\sigma_z=-1$ for both driving protocols (purple data points along the axis Im$\langle a\rangle$ in Fig.~\ref{fig7}), see Appendix~\ref{app:fig7}. 

Found the suitable parameters, we switch on the driving leading the qubit to the incoherent superposition $(|0\rangle\langle0|+|1\rangle\langle1|)/2$, such that  $\langle\sigma_z\rangle\sim0$. Figure \ref{fig7} shows that dispersive and longitudinal readout schemes exhibit a radically different behavior. Dispersive readout  mainly leads to a rotation of the pointer state as expected from a shift in the resonator frequency (blue signal). Parametric longitudinal readout moves the pointer state back to $(I,Q)\sim0$ (red signal). These outcomes are consistent with Eqs.~\eqref{Eq:a_par} and \eqref{Eq:a_disp}.

\section{Conclusion} 

We employed a triple quantum dot system to demonstrate how longitudinal coupling can be leveraged to perform site-selective readout in a quantum dot array. This is particularly relevant for large arrays of dots, for which crossbar networks have recently emerged as a promising approach \cite{li_crossbar_2018,Borsoi2023} for control and readout. It has been proposed to connect a resonator to each line of the array, with each line coupled to the individual dots. By applying drives to the columns of the array, one can select the readout location. This eliminates the need for mixing multiple tones \cite{john_bichromatic_2024}, which require more connections and introduce spurious harmonics.

By focusing on a single charge qubit, we were able to measure the coupling strength with excellent agreement with theoretical predictions. A key feature of longitudinal coupling is its linear (parametric) dependence on $A_q$, allowing, in principle, an increase in the measured signal. It is expected to saturate for strong drives $A_q > \omega_q$. Interestingly, the linear regime extends beyond this limit at high $\omega_q$ (Fig. ~\ref{fig5}), facilitating charge readout in the  ``ultra-dispersive" regime\cite{harpt_ultra-dispersive_2024}, where the dispersive coupling is strongly reduced. 

Charge qubits are however usually not coherent systems. Understanding the key features of the longitudinal coupling between a charge qubit and an RF resonator represents only an important step toward realizing the longitudinal coupling of hole spins\cite{yu_strong_2023} or singlet-triplet qubits\cite{bottcher_parametric_2022} to microwave cavities via spin-orbit interaction. Such advancements hold great potential for fast and QND readout, but also for enabling long-range spin-spin interactions in future multi-qubit architectures.

\section{Acknowledgments}
This research has been supported by the European Union’s Horizon 2020 research and innovation programm under grant agreement nos. 951852 (QLSI project), 810504 (ERC project QuCube) and 759388 (ERC project LONGSPIN), and by the French National Research Agency (ANR) through the PEPR PRESQUILE (ANR-22-PETQ-0003). V. C. acknowledge support from the Program QuantForm-UGA n° ANR-21-CMAQ-0003 France 2030 and by the LabEx LANEF n° ANR-10-LABX-51-01. A. C. and M. F. acknowledge support from EPiQ ANR-22-PETQ-0007 part of Plan France 2030C. B.B. acknowledges support from ANR-23-CPJ1-0033-01.

\section{Author contribution}
V.C. performed the measurements with input from S.Z. and R.M. V.C. analyzed the data with input from S.Z., \'{E}.D. and R.M. A.C. Y.-M.N. and M.F. helped develop the theoretical framework. B.Be. and H.N. were responsible for the fabrication of the device. X.J., V.S., B.Br., and C.W. participate in the global experimental effort. V.C., S.Z., M.F., S.D.F and R.M. co-wrote the manuscript with inputs from all the authors.

\appendix

\section{Resonator}\label{app:resonator}
\begin{figure}[h]
    \centering
    \includegraphics[width = \linewidth]{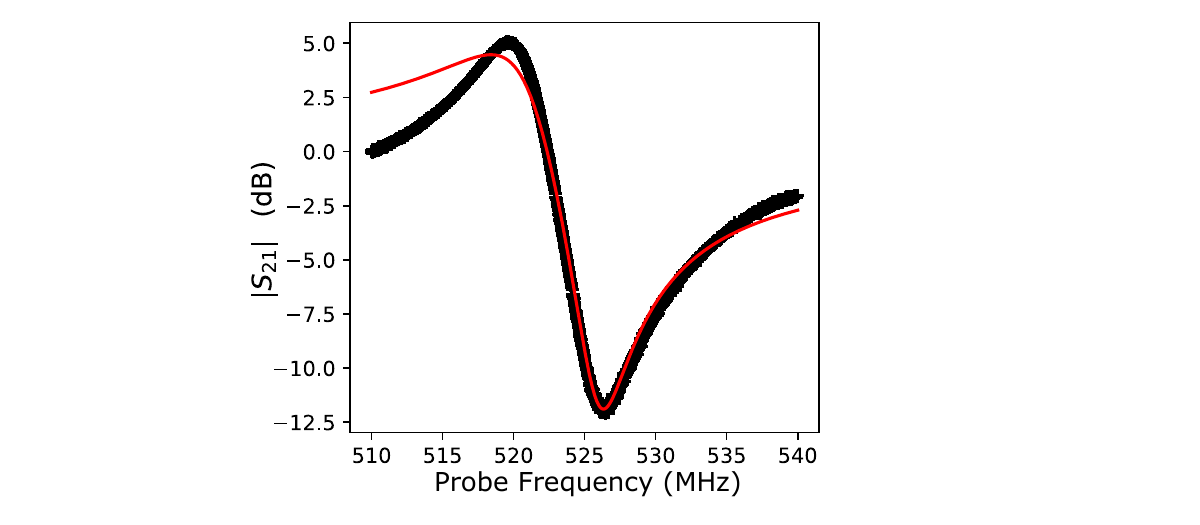}
    \caption{Resonance measured in transmission and fit to Eq.~\eqref{eq:hanger}.}
    \label{supp1}
\end{figure}
The resonator used in this work consists in surface mount copper inductor (with $210 \pm 10$ nH inductance) capacitively coupled to a transmission line, and galvanically coupled to the gate G2 of the device. To extract the resonator characteristics, the transmission amplitude  $ S_{21, \text{dB}}$ is fitted to the following equation: 
\begin{equation}
    S_{21, \rm dB} = 20 \log \left ( \left |1+\frac{Q_ie^{i\phi}}{Q_c\left ( 1+ 2iQ_i\left ( \frac{f-f_r}{f_r}\right )\right ) } \right | \right ) +S_0  \label{eq:hanger}
\end{equation}
 $Q_i$ is the internal quality factor, $Q_c$ is the external quality factor, $f_r$ is the resonance frequency, and $S_0$ is the background. The fit yields: 

\begin{align*}
    f_r&= 525.99 \pm 0.01 \text{ MHz}\\ 
    Q_i&= 164.3 \pm 0.6 \\ 
    Q_c&= 49.2 \pm 0.2 \\ 
\end{align*}

Assuming
\begin{align}
    f_r &= \frac{1}{2\pi \sqrt{LC}} \\
    Z_r &= \sqrt{\frac{L}{C}} \\
    \frac{\kappa}{2\pi} &= f_r\left (\frac{1}{Q_i}+\frac{1}{Q_c}\right ) 
\end{align}

We can extract the parallel capacitance $C$ the characteristic impedance $Z_r$ and the decay rate $\kappa$ of the resonator:

\begin{align*}
    C &= 436 \pm 22 \text{ fF}\\ 
    Z_r&= 694 \pm 25 \text{ }\ohm \\
    \kappa/2\pi&= 13.88 \pm 0.04 \text{ MHz}
\end{align*}

Knowing the characteristic impedance allows for estimating the zero point voltage fluctuation of the resonator, and hence the charge photon coupling.
\begin{equation}
    g_0/2\pi = \alpha \frac{e V_{ZPF}}{2h} =\alpha e f_r \sqrt{\frac{Z_r}{2h}}
\end{equation}
We find $g_0/2\pi = 25.0 \pm 1.3 $ MHz. Using $\alpha = 0.41\pm0.02$ (that we measure independently using non-adiabatic Landau-Zener-Stückelberg interferences\cite{dupont-ferrier_coherent_2013}) leads to the red prediction in Fig.~\ref{fig4}a of the dispersive shift $\chi$. 

We can estimate the photon number in the resonator with a model derived from an input-output theory for a given input power $P_{in}$ in the feedline
\begin{equation}
    \langle a^\dagger a \rangle = Q_c\left(\frac{Q_i}{Q_i+Q_c}\right )^2 \frac{P_{in}}{\hbar \omega_r^2}
    \label{nphoton}
\end{equation}

This allows for calibrating the $(I,Q)$ plane in $($Re$(\langle a \rangle ), $ Im$(\langle a \rangle))$. For an applied single tone to the cavity of known amplitude $A_R$, on one side we estimate the number of photons in the resonator using Eq.~\eqref{nphoton}. On the other side we measure the quadrature $I$ and $Q$ of the field at the output of the feedline (after the amplification chain). This gives us a calibration of the number of photons for a measured $I$, $Q$, that does not require to calibrate the exact gain of the amplification chain.  
\begin{figure}
    \centering
    \includegraphics[width =\linewidth]{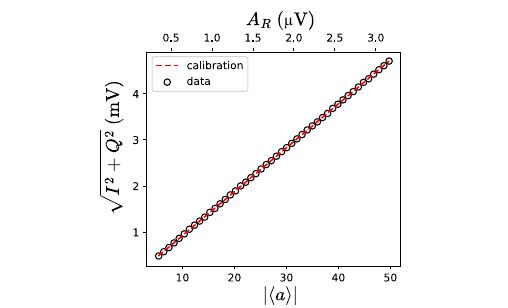}
    \caption{Calibration of the photon number with respect to the amplitude of the output field.}
    \label{supp2}
\end{figure}

\section{Charge configuration}\label{app:charge}
To form the charge qubit, we first load a few charges in a double QD by applying a negative voltage to G4 and G3, and bring back G4 back to zero voltage to separate these from the lead. Decreasing G2 now enable to move charges from G3 to G2, creating interdot transition. Provided their coupling is in a range where dispersive interaction is not negligible, we can measure these interdot transitions by measuring the phase shift of the transmitted signal through the transmission line and count them. By playing on the gate B3, we can tune the coupling to make appear all the transitions. We can repeat the loading sequence until we find just one charge in the double QD. 

\begin{figure}
    \centering
    \includegraphics[width = \linewidth]{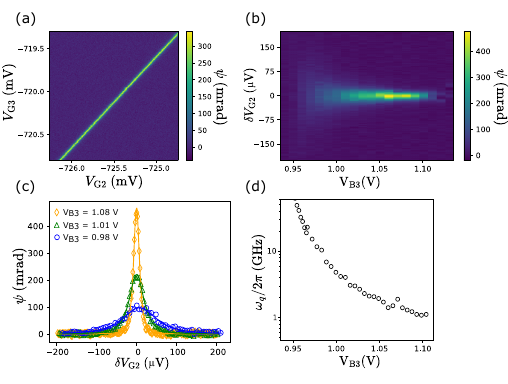}
    \caption{\textbf{Charge configuration. a.} Interdot transition in the (G2, G3) subspace. \textbf{b.} Interdot transition in the (G2, B3) subspace. \textbf{c.} Interdot transition as a function of $V_{G2}$ for three different values of $V_{B3}$ and fits to Eq.~\ref{eq:fit_interdot}. \textbf{d.} Qubit energy as a function of barrier gate voltage.}
    \label{fig:supp_charge}
\end{figure}

In this situation the charge stability diagram (see Fig.~\ref{fig:supp_charge}a) in the (G2, G3) subspace only shows an infinite interdot transition as a 45° degree line, since we separated the hole from the lead. We can tune the tunnel coupling of the the double QD by changing the voltage on gate B3. We report the evolution of the interdot transition in  Fig.~\ref{fig:supp_charge}b, as well as cuts at fixed B3 voltages in  Fig.~\ref{fig:supp_charge}c. While increasing $V_{B3}$ we increase the height of the potential barrier, decreasing the tunnel coupling. Therefore the interdot transition increases in amplitude and gets narrower as expected by the theory. We can indeed fit the interdot transition to the model: 

\begin{equation}
    \psi = A\times \frac{2t^2}{(\varepsilon^2+4t^2)^{3/2}}\tanh\left(  \frac{\left ( \varepsilon^2+4t^2\right )^{1/2}}{2k_BT}\right)  \label{eq:fit_interdot}
\end{equation}
 where $\psi$ is the phase of the output field, $A$ is a constant, $\varepsilon$ the detuning, $T$ the temperature and $t$ the tunnel coupling.  In doing so, we can extract the tunnel coupling. Figure \ref{fig:supp_charge}c. shows that the model agrees with the data, and \ref{fig:supp_charge}d. shows the evolution of the charge qubit energy $\omega_q = 2t$ with the barrier gate voltage.

\section{Field cancellation}\label{app:cancellation}
\begin{figure}
    \centering
    \includegraphics[width = \linewidth]{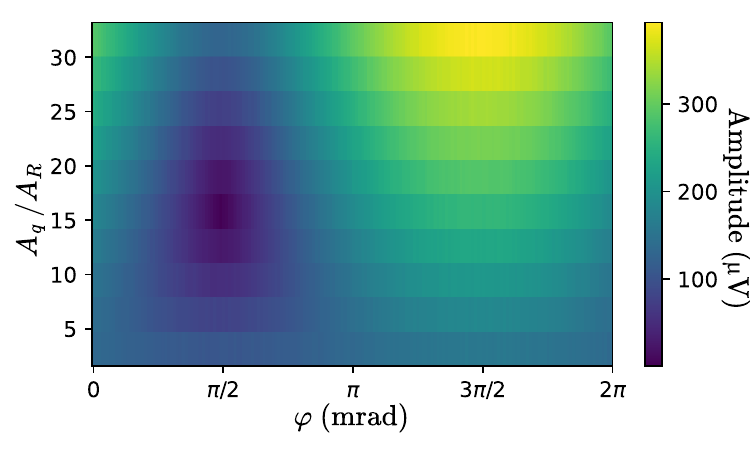}
    \caption{\textbf{Calibration of the cancelling tone.} We measure the amplitude of the field at the output of the transmission line. We sweep as parameters $\varphi$ and $A_q$.  We notice that for a phase difference $\varphi \sim 3\pi/2$ the amplitude of the field always increases while increasing $A_q$, indicating constructive interferences between $A_R$ and $A_q$. For $\varphi \sim \pi /2 $, the field finds a minimum value to 0 for a specific ratio $A_q/A_R \sim 15$. We can then calibrate $A_R$ and $\varphi$ to this values to compensate for the leakage of $A_q$ in the resonator.}
    \label{fig:supp_cancel}
\end{figure}

To cancel out the leakage of the longitudinal modulation into the transmission line, we cancel the coupling between the resonator and the qubit by increasing the double QD detuning $\varepsilon \gg \omega_q$. We then send the two tones: $A_q \cos{\omega_rt}$ to the qubit and $ A_R\cos(\omega_rt+ \varphi)$ in the transmission line, and measure the output of the transmission line. When the leakage of $A_q$ interferes destructively with $A_R$, the output amplitude must cancels out. We report the output amplitude as a function of the amplitude ratio and phase shift between the two tones in Fig.~\ref{fig:supp_cancel}. This allows for calibrating the cancelling tone. 

\section{Experimental protocol of Fig.  7}\label{app:fig7}

\begin{figure}
    \centering
    \includegraphics[width=\linewidth]{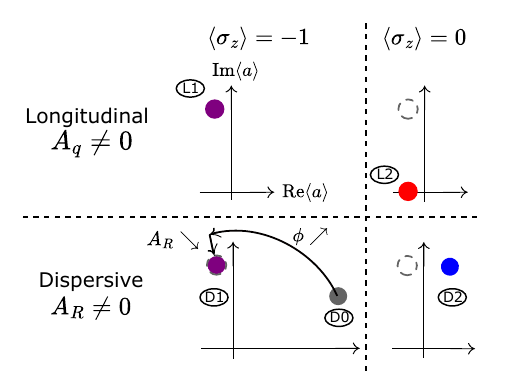}
    \caption{Experimental protocol for probing the conditional displacement. See text for step description}
    \label{fig:enter-label}
\end{figure}

We propose a visual experiment to probe the conditional displacement of the longitudinal readout and to differentiate it from dispersive read-out. The intrinsic idea is to perform two measurements, that starts from a same initial state, but that evolves towards different final states under a same transformation. The protocol is the following:
\begin{itemize}
\item[L1:] We prepare the qubit in its ground state, and activate the longitudinal interaction by a modulation $A_q \cos \omega_r t$ with the proper cancellation. One measures a resonator state $\langle a \rangle_{\parallel,\langle \sigma_z \rangle=-1} $.

\item[L2:] We drive the qubit to an incoherent mixture of ground and excited states, and the resonator is then displaced to a state $\langle a \rangle_{\parallel, \langle \sigma_z \rangle=0} $.

\item[D0:] We prepare now the qubit in its ground state, and activate only the dispersive interaction, by populating the resonator with an external drive $A_R\cos (\omega_rt+\phi)$ on the resonator (Note that $A_q = 0$).

\item[D1:] We prepare the resonator in the same state than during L1 by tuning $A_R$ and $\phi$ in such a way that the resonator steady state reads $\langle a \rangle_{\text{disp},\langle \sigma_z \rangle=-1}= \langle a \rangle_{\parallel,\langle \sigma_z \rangle=-1}$. 
\item[D2:] We drive again the qubit to an incoherent mixture, and the resonator reads  $\langle a \rangle_{\text{disp},\langle \sigma_z \rangle=0}$.

\end{itemize}

\vfill

\end{document}